\documentclass[prb,twocolumn]{revtex4-1}
\usepackage{graphicx}
\usepackage{hyperref}
\usepackage{amssymb}
\usepackage{dcolumn}
\usepackage{float}
\usepackage{bm}
\usepackage{multirow}
\usepackage{booktabs}
\usepackage[T1]{fontenc}
\usepackage[utf8]{inputenc}
\usepackage{lmodern}
\usepackage{color}
\usepackage{makecell}
\usepackage{tikz}
\usepackage{pgfplots}
\usepackage{mathtools}
\usepackage{amsmath}

\newcommand{\mathleft}{\@fleqntrue\@mathmargin0pt}

\newcolumntype{M}[1]{>{\centering\arraybackslash}m{#1}}

\newcommand{\tens}{%
	\mathbin{\mathop{\otimes}}%
}
\DeclareMathAlphabet{\bi}{OML}{cmm}{b}{it}

\def\be{\begin{equation}}
\def\ee{\end{equation}}
\def\bearr{\begin{eqnarray}}
\def\eearr{\end{eqnarray}}

\def\d{\dow narrow}

\begin{document}
	
\title{Evolution of Majorona zero-energy edge states in a $T^2 = -1$ symmetry protected 1D topological superconductor with dominant spin-orbit coupling
}
\bigskip
\author{Alestin Mawrie}
\normalsize
\affiliation{Department of Physics, School of Physical Sciences, Mizoram University, Aizawl, Aizawl-796004, India}
\date{\today}
\begin{abstract}
We consider a 1D topological superconductor (TSC) constructed by coupling a pair of Kitaev's Majorana chains with opposite spin configurations. Such a 1D lattice model is known to be protected by a $T^2 = -1$ time-reversal symmetry. Furthermore, we consider a modeled Rashba spin-orbit coupling on such a system of $T^2=-1$ time-reversal symmetric TSC. The Rashba spin-orbit coupling
together with the chemical potential engineered the phase transitions of the edge states in the system and consequently the number of Majorona's zero-energy edge modes (MZM's) emerging at the edge of the coupled chains. Correspondingly, the topological nature of the system is described by a phase diagram consisting of three different phases. The three phases are characterized by a topological winding number, $\mathcal{W}=1$, $2$ (with one and two MZM's: topological phases) and  $\mathcal{W}=0$ (devoid of any MZM: trivial insulating phase).
\end{abstract}

\email{mzut251@mzu.edu.in}

\maketitle
An increasing effort in developing the future quantum computer technology has led to full swing research on low-input power integrating circuits. 
Owing to their conducting edge states, topological insulators are one of those candidates that offer most of all the desirable demands. Ever since it was introduced by Haldane\cite{Haldane}, it has been at the center of condensed matter research.
Topological insulators have chiral edge states which essentially do not require any external magnetic fields\cite{CES1,CES2} for them to exhibit the quantum Hall effect \cite{CES3}, thus often called anomalous quantum Hall effect. With no such requirements of an external magnetic field, topological insulators are very much sought after when called upon for a future low-power-consuming integrated circuit application. 

One of many premier discoveries in condensed matter physics has been that of the Majorana fermions in a topological superconductor (TSC's) (a distinct class of superconductors).  
Majorana fermion is being viewed as a ``half-fermion'' that obeys non-Abelian statistics.\cite{sato}. In a TSC, these fermions become localized at the TSC's edges to form zero-energy bound state modes called Majorana zero modes (MZM's)\cite{R1,R2,R3,R4,R5}.  Experimental detections of MZM's are done by transferring topological properties on a conventional superconductor and alternately, by making a contact between a superconductor and a topological insulator\cite{Trang,tsc3,tsc4,Ex1,Ex2,Ex3,Ex4,Ex5,Ex6}. 
Theoretically, the most accepted model that formulates the MZM's was proposed by Kitaev\cite{kitaev}. Kitaev  proposed a TSC in a 1-dimension chain consisting of an arrangement of spinless electrons with nearest-neighbor hopping energy `$t$' and superconducting energy gap `$\Delta$'. If the chemical potential, `$\mu$', is engineered in such a way that $|\mu|\le |2t|$ (with $\Delta\ne 0$), the chain enters a topological phase and hosts unpaired MZM's at each of its ends. \cite{kitaev,Leumer}
\begin{figure}[b]
\includegraphics[width=88mm,height=40.5mm]{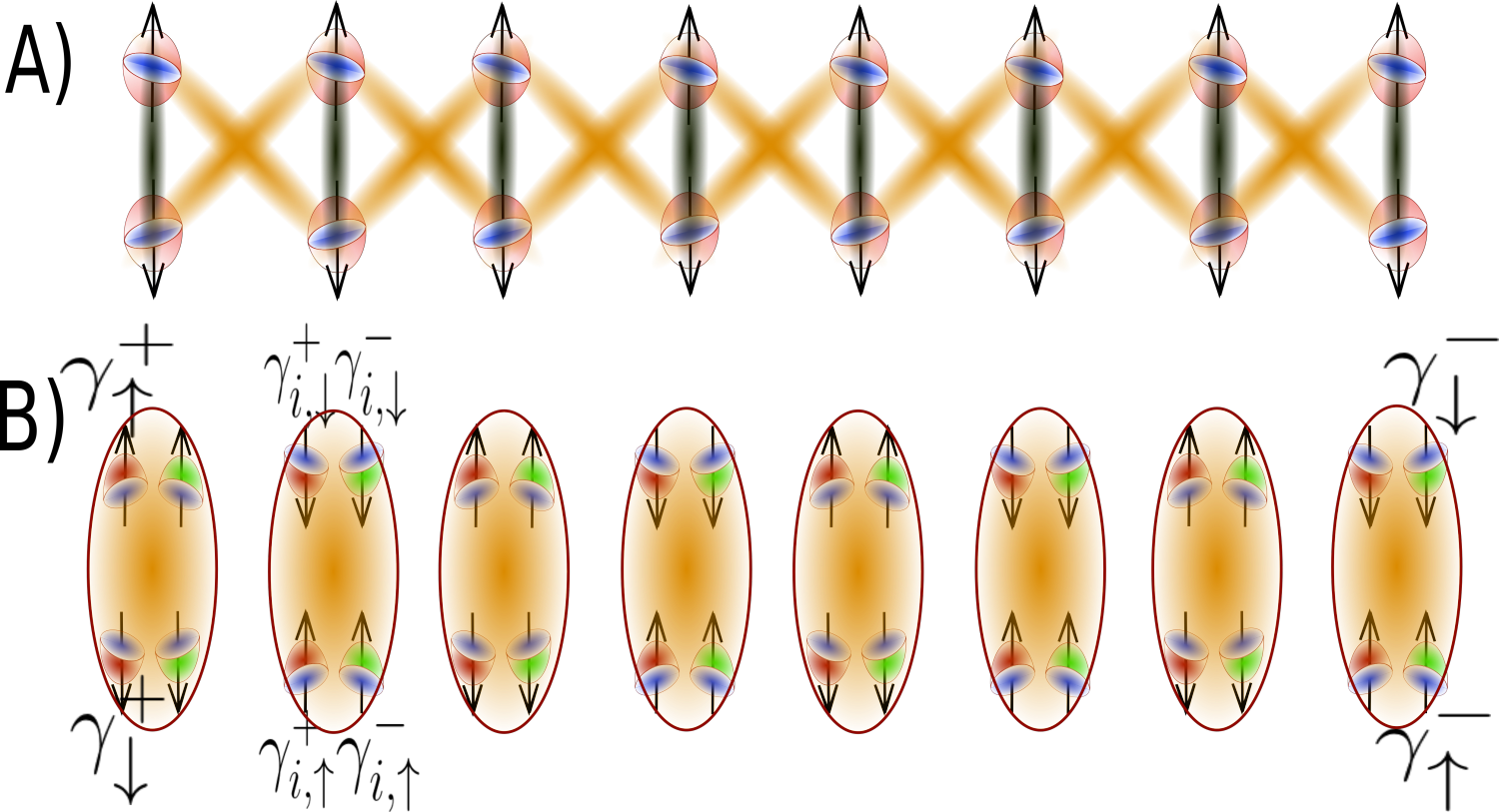}
\caption{A time reversal symmetric $T^2=-1$ protected 1D TSC. A) A coupling of two Majorana chains with opposite spin configurations, emphasising the Rashba spin-orbit couplings. B) 
The figure depicts the Majorana spinor pair of opposite spins at each ends of the coupled Kitaev's Majorana chains, $\gamma_\uparrow^+=\gamma_{1,\uparrow}^+$, $\gamma_\downarrow^+=\gamma_{1,\downarrow}^+$ on the left and $\gamma_\uparrow^-=\gamma_{N,\uparrow}^-$, $\gamma_\downarrow^-=\gamma_{N,\downarrow}^-$ on the right.
	}
	\label{Fig2}
\end{figure}

In this paper, we consider a special class of a 1D TSC protected by the $T^2=-1$ time-reversal symmetry\cite{T21}. The coupling of two Kitaev's Majorana chains with opposite spin configurations (as shown in Fig. \ref{Fig2}) results in this special symmetry-protected topological phase\cite{spt}. 
In such an arrangement, each physical site can be considered to consist of four Majorana modes (see Fig. [\ref{Fig2} (B)]). In this paper, we will show the different conditions on the system parameters (also with the inclusion of spin-orbit coupling effect) by which the chain will enter different topological phases and correspondingly, the phase transitions of the edge states. 
It is to be noted that the dangling zero-energy Majorana spinons at both ends of the chain become protected by the time-reversal symmetry\cite{T21} that reads as 
\begin{eqnarray}\label{l1}
\left.
\begin{aligned}
Td_{i\uparrow}T^{-1}=-d_{i\downarrow},\hspace{0.5cm}Td_{i\downarrow}T^{-1}=d_{i\uparrow}\\
Td_{i\uparrow}^\dagger T^{-1}=-d_{i\downarrow}^\dagger,\hspace{0.5cm}Td_{i\downarrow}^\dagger T^{-1}=d_{i\uparrow}^\dagger.
\end{aligned}
\right\}
\end{eqnarray}

The complex fermioninic operators 
$d_{i\sigma}$ and $d_{i\sigma}^\dagger$ (with $\sigma=\uparrow,\downarrow$ being the spin of the fermion) can be written in terms of the two Majorana operators $\gamma_{i\sigma}^+$ and $\gamma_{i\sigma}^-$ as 
\begin{eqnarray}
\left.
\begin{aligned}d_{i\sigma}=\frac{1}{2}(\gamma_{i\sigma}^++i\gamma_{i\sigma}^-)\\
d_{i\sigma}^\dagger=\frac{1}{2}(\gamma_{i\sigma}^+-i\gamma_{i\sigma}^-)
\end{aligned}
\right\}
\end{eqnarray}

Similar to the spinless Majorana fermions\cite{kitaev}, the above Majorana operators $\gamma_{i\sigma}^+$ and $\gamma_{i\sigma}^-$ have the anti-commutation relations 
$\{\gamma_{i\sigma}^\pm,\gamma_{j\sigma^\prime}^\mp\}=0$ and $\{\gamma_{i\sigma}^\pm,\gamma_{j\sigma^\prime}^\pm\}=2\delta_{ij}\delta_{\sigma,\sigma^\prime}$. From Eq. (\ref{l1}), it is easy to see that the time-reversal symmetry has a projective representation of  $T^2 = -1$ for the fermion parity odd basis and  $T^2 = 1$ for fermion parity the even basis. To construct a total symmetry group for this system 
of spin-full fermions we need four group elements \{$I,T,T^2,T^3$\} with $T^4 = 1$, which is a $\mathbb{Z}_4$ group, an extension of the $\mathbb{Z}_2$ fermion parity symmetry group.\cite{T21}

Nanowires such as InSb, have strong spin-orbit coupling with a large \textit{g} factor\cite{InSb1,InSb2,Chang,InSb3,InSb4,InSb5,InSb6,InSb7}. 
The spin-orbit coupling dictates the size of the topological phases in a TSC\cite{dasS}, thus devices with strong spin-orbit coupling play a crucial role in studying the Majorana  fermions.
In this paper, we modeled a possibility of flipping the spin of the neighboring fermions by means of the Rashba spin-orbit coupling\cite{Chang,InSb3} (depicted in Fig. [\ref{Fig2}]). 
The Hamiltonian of this system of spinful Kitaev Majorana chains with Rashba spin-orbit coupling\cite{soc,theorySupport} is 
\begin{eqnarray}\label{HamilR}
	&H=-\mu\sum_{s,j}^Nd_{j,\sigma}^\dagger d_{j,\sigma}+ \sum_{\sigma\ne \sigma^\prime,j=1}^{N-1}[-td_{j,\sigma}^\dagger d_{j+1,\sigma}\nonumber\\&+\Delta d_{j,\sigma}^\dagger d_{j+1,\sigma}^\dagger+\alpha d_{j,\sigma}^\dagger d_{j+1,\sigma^\prime} -\alpha_z d_{j,\sigma}^\dagger d_{j,\sigma^\prime}+\text{h. c.}].
\end{eqnarray}
Here the four parameters that characterized this Hamiltonian are 
the on-site chemical potential, `$-\mu$', the fermion hopping energy, `$t$', the superconducting pairing energy, `$\Delta$', the Rashba spin-orbit coupling strength `$\alpha$' and the Zeeman's type energy, `$\alpha_z$'. 
To allow the pairing of electrons with the same spin species, we considered a TSC with a superconducting supply that are domain walls of 1D $p$-wave superconductors \cite{pwave1,pwave2,pwave3,kitaev,theorySupport}.

\begin{figure}[b]
	\includegraphics[width=65mm,height=42.5mm]{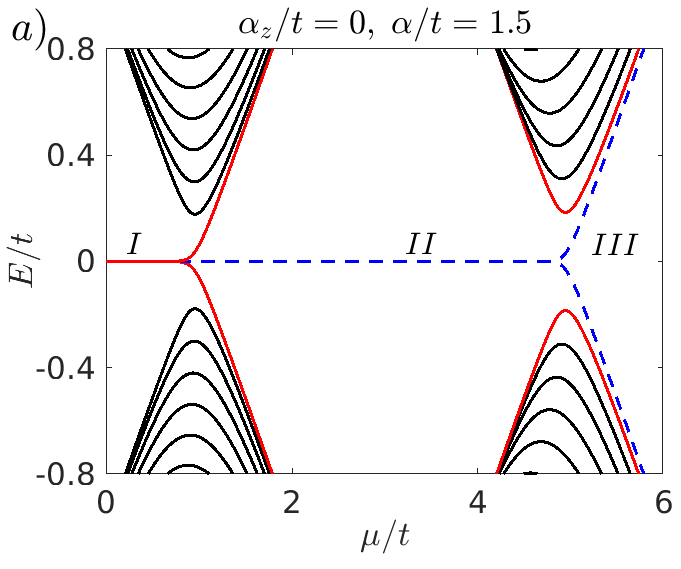}
	\includegraphics[width=88mm,height=38.5mm]{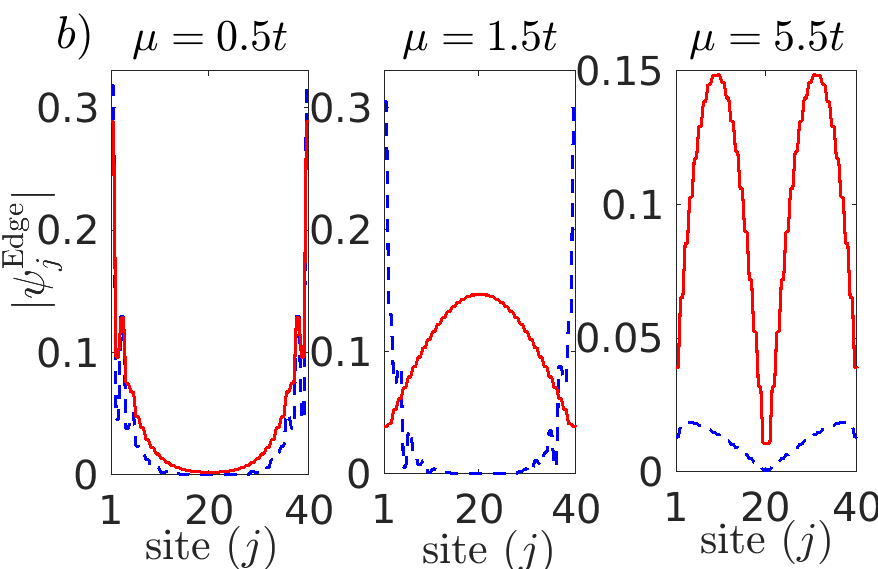}
	\caption{(a): Band diagram of the first 7 bands (counting from the first edge state) for a total number of sites $N=40$. The two edge states are shown in solid red and dashed blue color, respectively. (b): Plot of the edge states wave-function with respect to the site number [for those bands shown in red and blue of Fig. (a)] for different values of $\mu$, setting $\alpha=1.5t$.
	}
	\label{Fig4}
\end{figure}

Taking the superconducting pairing, $\Delta=t$, the energy bands using the above real space Hamiltonian in Eq. (\ref{HamilR}) is shown in Fig. (\ref{Fig4}). On increasing $\mu$, the two edge states undergo a phase transition from their topological phases (regimes \textit{I} and \textit{II}) to a trivial insulating phase (regime \textit{III}). There are two and one MZM's in the regimes \textit{I} and \textit{II}, respectively. Whereas, regime \textit{III} is devoid of the MZM's. 
To have a clear insight of the nature of these MZM's, we plot the amplitude of the wavefunction, `$|\psi_j|$' versus the site number, `$j$' in Fig. \ref{Fig4} (b),  for the first two states in $E/t>0$ (depicted in solid red color and dashed blue color).
The MZM's in regimes \textit{I} and \textit{II} are localized towards the end of the chain at $j=1$ and $j=N$ indicating that the MZM's are contributions from those states at the two edges of the chain. In regime \textit{III}, the wave-function of the two edge states has significant contributions from those sites at the bulk of the chain.

For analyzing the bulk spectrum, we write down the fermionic operator $d_{j,\sigma}$ in Eq. (\ref{HamilR}) as
\begin{eqnarray}
d_{j,\sigma}=\frac{1}{\sqrt{N}}\sum_{k}d_{k,\sigma}e^{ik j}
\end{eqnarray}
where we have taken a unit lattice constant and the momentum, `$k$' is restricted within the first Brillouin zone, i.e. $k\in [-\pi,\pi]$.
Using the above equation for the fermionic operator, the Hamiltonian  in Equation (\ref{HamilR}), can now be written as
\begin{eqnarray}\label{k2}
H&=&-\mu\sum_{s,k}d_{k,\sigma}^\dagger d_{k,\sigma}+\sum_{s\ne s^\prime,k}[-t e^{ik}d_{k,\sigma}^\dagger d_{k,\sigma}\nonumber\\&+&\Delta e^{-ik} d_{-k,\sigma}^\dagger d_{k,\sigma}^\dagger-(\alpha_z-\alpha e^{ik }) d_{k,\sigma}^\dagger d_{k,\sigma^\prime}+\text{h. c.}].
\end{eqnarray}
To find the $k$-space Bogoliubov-de Gennes (BdG)
Hamiltonian, we choose the $k$-space Nambu spinor as $\psi_{k,\uparrow,\downarrow}=\begin{pmatrix}
d_{k,\uparrow} & d_{k,\downarrow} &d_{-k,\uparrow}^\dagger& d_{-k,\downarrow}^\dagger
\end{pmatrix}^\prime$(where $^\prime$ denotes the transpose) such that the Hamiltonian in Eq. (\ref{k2}) can be written as

\begin{widetext}
\begin{eqnarray}\label{Hk}
	H&=&\psi_{k,\uparrow,\downarrow}^\dagger \mathcal{H}(k) \psi_{k,\uparrow,\downarrow}\nonumber\\&=&\begin{pmatrix}
d_{k,\uparrow}^\dagger & d_{k,\downarrow}^\dagger &d_{-k,\uparrow}&d_{-k,\downarrow}
	\end{pmatrix} \begin{pmatrix}
		-\mu-2t\cos k & -\alpha_z+2\alpha \cos k &2i \Delta \sin k  &0 \\
		-\alpha_z+2\alpha \cos k & -\mu-2t\cos k & 0 & 2i \Delta\sin k \\
		-2i \Delta\sin k & 0& \mu+2t\cos k &  \alpha_z-2\alpha \cos k\\
		0& -2i \Delta\sin k & \alpha_z-2\alpha \cos k & \mu+2t\cos k
	\end{pmatrix} \begin{pmatrix}
d_{k,\uparrow} \\ d_{k,\downarrow} \\d_{-k,\uparrow}^\dagger\\ d_{-k,\downarrow}^\dagger
\end{pmatrix},
\end{eqnarray}
The $4\times 4$ Bogoliubov-de Gennes (BdG) Hamiltonian, $\mathcal{H}(k)$
is easily diagonalized to yield a 4-band energy spectrum given by  
\begin{eqnarray}
E_s^\lambda(k)=s \sqrt{(\mu+2t\cos k)^2+(\mu+2\lambda\alpha\cos k)^2+(\mu-\lambda\alpha_z)^2+8\lambda \alpha t\cos^2k+4\Delta^2\sin^2k-2(\mu^2+2\alpha_z(\alpha+\lambda t)\cos k)}.
\end{eqnarray}
\end{widetext}
Here, $s,\lambda=\pm$. In the limit of $\alpha\rightarrow 0$ and $\alpha_z\rightarrow 0$, the above result reduced to a system of two Kitaev chains that hosted two degenerated unpaired MZM's when $|\mu|<2|t|$.\cite{Leumer}   
As expected, the BdG Hamiltonian in Eq. (\ref{Hk}) should exhibit a time-reversal symmetry $\mathcal{T}^2 \mathcal{H}^\ast(k)(\mathcal{T}^{-1})^2=\mathcal{H}(-k)$, with $\mathcal{T}=\sigma_z\tens \sigma_z$. Also, it has an electron-hole symmetry $\xi \mathcal{H}^\ast(k)\xi^{-1}=-\mathcal{H}(-k)$, 
with $\xi=\sigma_x\tens\sigma_0$. Here ($\sigma_0,\sigma_x,\sigma_y,\sigma_z$) are the Pauli's matrices. 
To reduce the BdG Hamiltonian into an off-diagonal matrix block, we introduce a unitary operator, `$\mathcal{U}$', defined as
\begin{eqnarray}
\mathcal{U}=\frac{1}{\sqrt{2}}\begin{pmatrix}
\sigma_x & \sigma_x\\
-\sigma_y & \sigma_y 
\end{pmatrix}.\nonumber
\end{eqnarray}
On doing the unitary transformation $\mathcal{U}^\dagger \mathcal{H}(k)\mathcal{U}$, we obtained a transformed BdG Hamiltonian
\begin{eqnarray}\label{bdg}
\mathcal{U}^\dagger \mathcal{H}(k)\mathcal{U}=\begin{pmatrix}
	\mathcal{O} & Q(k)\\ Q^\dagger(k) & \mathcal{O}
\end{pmatrix}.
\end{eqnarray}
\begin{figure}[t]
	\includegraphics[width=88mm,height=65.5mm]{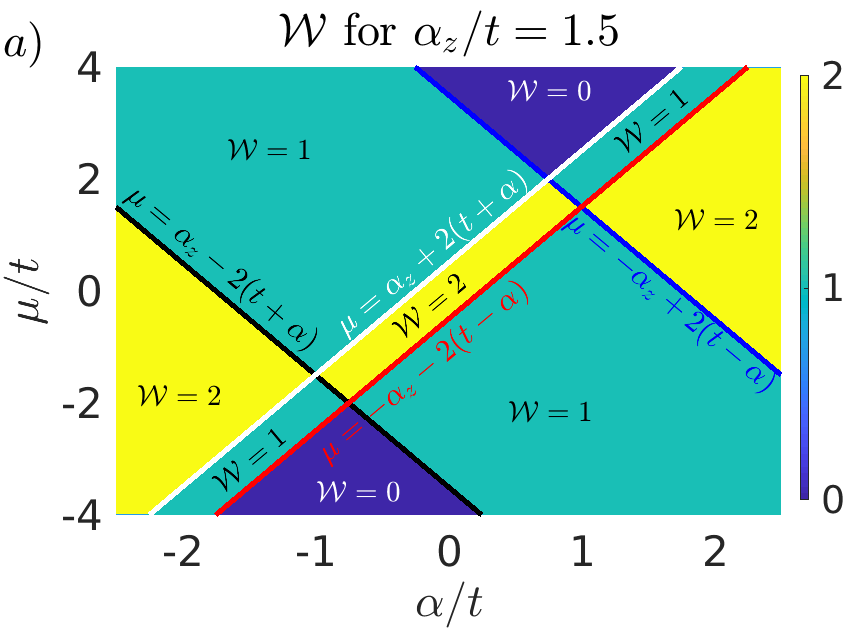}
	\includegraphics[width=42mm,height=35mm]{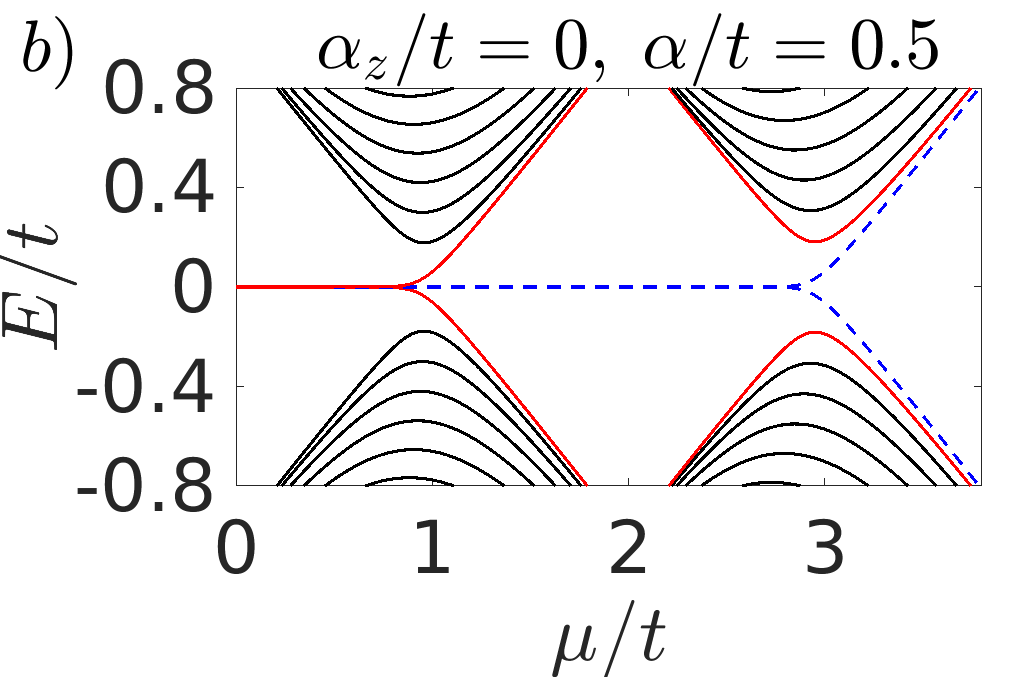}
	\includegraphics[width=42mm,height=35mm]{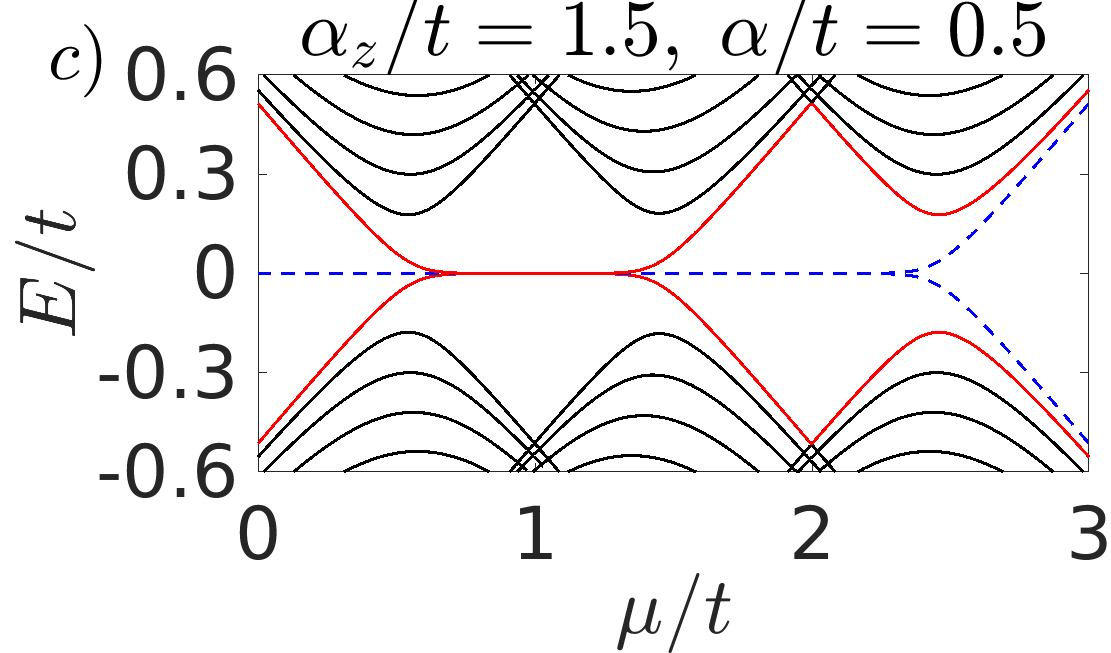}
	\caption{a) Plots of the winding number as a function of $\alpha/t$ and $\mu/t$ for $\alpha_z/t=1.5$. The different topological regimes are dictated by the condition $|\mu\mp \alpha_z |<2|t\pm \alpha|$. b) Plots of the few bands for $\alpha/t=0.5$ and $\alpha/t=1.5$ (from right to left), taking $N=40$.}
	\label{Fig5}
\end{figure}
Here $\mathcal{O}$ is a $2\times 2$ null matrix and $Q(k)$ is given by 
\begin{eqnarray}\label{Qk}
Q(k)&=&i(\mu+2t\cos k)\sigma_z-2\Delta\sin k\sigma_z\nonumber\\&+&(\alpha_z-2\alpha\cos k)\sigma_y.
\end{eqnarray}
We note that ${\rm det}[\mathcal{H}({k})] = {\rm det}[Q(k)] \cdot {\rm det}[(Q^\dagger({k})]$. For a band closing condition, ${\rm det}[\mathcal{H}({k})]=0$. We, thus, arrive at the two conditions for closing the gap which are given as:
\begin{eqnarray}\label{gs}
	\left.
	\begin{aligned}
	(\mu-\alpha_z+2(t+\alpha) \cos k)^2+4\Delta^2\sin^2k=0\\
		(\mu+\alpha_z+2(t-\alpha) \cos k)^2+4\Delta^2\sin^2k=0
	\end{aligned}
	\right\}.
\end{eqnarray}
The above two conditions basically mark the boundaries for the appearance of the spin-up and spin-down MZM's, respectively. 
Similar to the case of a single Majorana Kitaev's chian\cite{kitaev,Leumer}, we now have an appearance of a gap closing at $k=0,\text{ and }k=\pm\pi$, when the superconducting gap $\Delta\ne 0$ and the remaining parameters set to $\mu\mp\alpha_z= -2(t\pm \alpha)$ and $\mu\mp\alpha_z= 2(t\pm \alpha)$, respectively. The system of two Majorana Kitaev's chains enters any of the topological phases when the different parameters dictate the conditions $\Delta\ne 0$ and $|\mu\mp\alpha_z|< 2|t\pm \alpha|$. 
In order to understand the characterization of the different topological phases, we calculate the winding number for the BdG Hamiltonian in Eq. (\ref{bdg}) \cite{w3,w1,w2,w4}.
\begin{eqnarray}
	\mathcal{W}=\text{Im }\int_{-\pi}^{\pi}\frac{d}{dk}\{\ln [\det[Q(k)]]\}.
\end{eqnarray}

We show the variation of this number as a function of $\mu$ and $\alpha$ in Figure. [\ref{Fig5} (a)], for the cases of $\alpha_z=1.5 t$.
The boundaries separating the different topological phases are determined by the equation $|\mu\mp\alpha_z|= \pm 2|t\pm \alpha|$ obtained by the gap closing conditions in Eq. (\ref{gs}).
The winding number gives 
the number of MZM's at each edge of the chain 
\begin{eqnarray}
	\text{No's of MZM's}=\begin{cases}
		2,              & \text{if }
		\mathcal{W}=2\\
		1,& \text{if } \mathcal{W}=1\\
		0, & \text{if } \mathcal{W}=0
	\end{cases}
\end{eqnarray}
To demonstrate the different topological phases we plot the band diagram as a function of the chemical potential for different strengths of the two types of spin-orbit couplings in Fig. [\ref{Fig5} (b \& c)]. For better understanding, the bands coming from the two edge states are shown in dashed-blue and solid-red colors. In the band diagram shown in section Fig. [\ref{Fig5} (b)], the Zeeman type spin-orbit coupling is set to zero. In this case, two MZM's appear for $\mu/t\le 1$ at the end of the chain, thereafter, one of those modes starts to undergo a phase transition to an insulating state on increasing the chemical potential. On further increasing of the chemical potential, the system enters the trivial insulating phase (when both the two edge states have been transitioned to their insulating states). In
Fig. [\ref{Fig5} (c)], we plot the band diagram with the spin-orbit coupling parameters set to $\alpha_z=1.5t$, and $\alpha=0.5t$. One of the states (indicated by red color) enters a topological phase only in the range $0.5t\le\mu\le1.5t$ and remains insulating for $\mu$ not in this range. Whereas there is only one phase transition from the other edge state (shown in dashed blue color) in the range $0\le\mu/t\le3$.

In conclusion, we have studied the phase transitions of the edge states in a Rashba spin-orbit coupling dominated 1D topological superconductor (TSC). The 1D TSC we considered is protected by a $T^2=-1$ time-reversal symmetry and is constructed by coupling a pair of Kitaev's Majorana chains with opposite spin configurations. The Rashba spin-orbit coupling is capable of engineering the phase transitions of the edge states, subsequently, the number of MZM's at the edge of the 1D system. Depending on the strength of the Rashba spin-orbit coupling and the chemical potential, the system identified two topological phases (with topological winding number, $\mathcal{W}=1$ and $2$) and a trivial insulating phase (with a topological winding number $\mathcal{W}=0$). The two topological phases are characterized by one (for $\mathcal{W}=1$) and two (for $\mathcal{W}=2$) MZM's at the edge of the two coupled Kitaev's Majorana chains.

\textit{Acknowledgments}: This work is an outcome of
the Research work carried out under the DST-INSPIRE project DST/INSPIRE/04/2019/000642, Government of India.


\begin{thebibliography}{55}

\bibitem{Haldane}
F. D. M. Haldane, \textit{Model for a Quantum Hall Effect without Landau Levels: Condensed-Matter Realization of the ``Parity Anomaly''}, Phys. Rev. Lett. \textbf{61}, 2015 (1988).

\bibitem{CES1}
Xi-Rong Chen, Wei Chen, L. B. Shao, and D. Y. Xing, \textit{Engineering chiral edge states in two-dimensional topological insulator/ferromagnetic	insulator heterostructures}, Phys. Rev. B, \textbf{99}, 085417 (2019).

\bibitem{CES2}
D. J. Thouless, M. Kohmoto, M. P. Nightingale, and M. den
Nijs, \textit{Quantized Hall Conductance in a Two-Dimensional Periodic Potential}, Phys. Rev. Lett. \textbf{49}, 405 (1982).

\bibitem{CES3}
K. V. Klitzing, G. Dorda, and M. Pepper, \textit{New Method for High-Accuracy Determination of the Fine-Structure Constant Based on Quantized Hall Resistance}, Phys. Rev. Lett. \textbf{45},
494 (1980).

\bibitem{sato}
M. Sato, and Y. Ando, \textit{Topological superconductors: a review},
Rep. Prog. Phys. \textbf{80}, 076501 (2017).
	
\bibitem{R1}
 J. Alicea, \textit{New directions in the pursuit of Majorana fermions insolid state systems}, Rep. Prog. Phys. \textbf{75}, 076501(2012).
 
 \bibitem{R2}  M. Leijnse and K. Flensberg, \textit{Introduction to topological super-conductivity and Majorana fermions},
 Semicond. Sci. Techn. \textbf{27}, 124003 (2012).
 
 \bibitem{R3}  
 C. W. J. Beenakker, \textit{Search for Majorana fermions in supercon-ductors},
 Annu. Rev. Con. Mat. Phys. \textbf{4}, 113 (2013).
 
 \bibitem{R4}  S.  R.  Elliott  and  M.  Franz,  \textit{Majorana  fermions  in  nuclear, particle, and solid-state physics},
 Rev. Mod. Phys. \textbf{87}, 137(2015).
 
 \bibitem{R5}  C.   W.   J.   Beenakker,   
 \textit{Random-matrix   theory   of   Majoranafermions and topological superconductors}, Rev. Mod. Phys. \textbf{87}, 1037(2015).
 

 
\bibitem{Trang}
C. X. Trang, N. Shimamura, K. Nakayama, S. Souma, K. Sugawara, I. Watanabe, K. Yamauchi, T. Oguchi, K. Segawa, T. Takahashi, Yoichi Ando and T. Sato, \textit{Conversion of a conventional superconductor into a topological superconductor by topological proximity effect}, 
Nature Communications \textbf{11}, 159 (2020).

\bibitem{tsc3}
L. Fu, and C. L. Kane, \textit{Superconducting proximity effect and Majorana fermions
at the surface of a topological insulator}, Phys. Rev. Lett. \textbf{100}, 096407 (2008).

\bibitem{tsc4} 
M. Sato, and S. Fujimoto, \textit{Topological phases of noncentrosymmetric
superconductors: edge states, Majorana fermions, and non-Abelian statistics}
Phys. Rev. B \textbf{79}, 094504 (2009).

\bibitem{Ex1}
V. Mourik, K. Zuo, S. M. Frolov, S. R. Plissard, E. P. A. M.
Bakkers, and L. P. Kouwenhoven, \textit{Signatures of Majorana Fermions in Hybrid Superconductor-Semiconductor Nanowire Devices}, Science \textbf{336}, 1003 (2012).

\bibitem{Ex2} 
M. T. Deng, S. Vaitiekenas, E. B. Hansen, J. Danon M. Leijnse,
K. Flensberg, J. Nygard, P. Krogstrup, and C. M. Marcus, \textit{Majorana bound state in a coupled quantum-dot hybrid-nanowire system},
Science \textbf{354}, 6319 (2016).

\bibitem{Ex3} 
H. O. H. Churchill, V. Fatemi, K. Grove-Rasmussen, M. T. Deng,
P. Caroff, H. Q. Xu, and C. M. Marcus, \textit{Superconductor-nanowire devices from tunneling to the multichannel regime: Zero-bias oscillations and magnetoconductance crossover}, Phys. Rev. B \textbf{87}, 241401
(2013).

\bibitem{Ex4} A. Das, Y. Ronen, Y. Most, Y. Oreg, M. Heiblum, and H.
Shtrikman, \textit{Evidence of Majorana fermions in an Al - InAs nanowire topological superconductor}, Nat. Phys. \textbf{8}, 887 (2012).

\bibitem{Ex5} 
L. P. Rokhinson, X. Liu, and J. K. Furdyna, \textit{Observation of the fractional ac Josephson effect: the signature of Majorana particles}, Nat. Phys. \textbf{8}, 795
(2012).


\bibitem{Ex6} 
S. M. Albrecht, A. P. Higginbotham, M. Madsen, F. Kuemmeth,
T. S. Jespersen, J. Nygard, P. Krogstrup, and C. M. Marcus, \textit{Exponential Protection of Zero Modes in Majorana Islands},
Nature (London) \textbf{531}, 206 (2016).

\bibitem{kitaev}
A. Y. Kitaev, \textit{Unpaired Majorana fermions in quantum wires}, Phys.-Usp. \textbf{44} 131 (2001).


\bibitem{Leumer}
N. Leumer, M. Marganska, B. Muralidharan
and M. Grifoni, \textit{Exact eigenvectors and eigenvalues of the
finite Kitaev chain and its topological
properties}, J. Phys.: Condens. Matter \textbf{32}, 445502 (2020).
%
%
%
%
%
%
%

\bibitem{T21}
Zheng-Cheng Gu, \textit{Fractionalized time reversal, parity, and charge conjugation symmetry in a topological
superconductor: A possible origin of three generations of neutrinos and mass mixing}, Phys. Rev. Research. {\bf 2}, 033290 (2020).


\bibitem{spt}
Z. C. Gu and X. G. Wen, \textit{Tensor-entanglement-filtering renormalization approach and symmetry protected topological order},
Phys. Rev. B \textbf{80}, 155131 (2009).


\bibitem{InSb1}
S. R. Plissard, D. R. Slapak, M. A. Verheijen, M. Hocevar, G. W. G. Immink, I. van Weperen, S. Nadj-Perge, S. M. Frolov, L. P. Kouwenhoven, and Erik P. A. M. Bakkers, \textit{From InSb Nanowires to Nanocubes: Looking for the Sweet Spot}, Nano Lett. \textbf{12}, 1794 (2012).

\bibitem{InSb2}
H. A. Nilsson, P. Caroff, C. Thelander, M. Larsson, J. B. Wagner, L. Wernersson, L. Samuelson, and H. Q. Xu, \textit{Giant, Level-Dependent g Factors in InSb Nanowire Quantum Dots}, Nano Lett. \textbf{9}, 3151 (2009).

\bibitem{Chang}
X. W. Zhang and J. B. Xia, \textit{Rashba spin-orbit coupling in InSb nanowires under transverse electric field}, Phys. Rev. B. \textbf{74}, 075304 (2006).

\bibitem{InSb3}
I. van Weperen, B. Tarasinski, D. Eeltink, V. S. Pribiag, S. R. Plissard, E. P. A. M. Bakkers, L. P. Kouwenhoven and M. Wimmer, 
\textit{Spin-orbit interaction in InSb nanowires},
Phys. Rev. B \textbf{91}, 201413(R) (2015).

\bibitem{InSb4}
V.  Mourik,   K.  Zuo,   S.  M.  Frolov,   S.  R.  Plissard,E.  P.  A.  M.  Bakkers,  and  L.  P.  Kouwenhoven,  Science \textbf{336}, 1003 (2012).

\bibitem{InSb5}  
A. Das, Y. Ronen, Y. Most, Y. Oreg, M. Heiblum, andH. Shtrikman, Nat. Phys. \textbf{8}, 887 (2012).

\bibitem{InSb6} 
M. T Deng, C. L. Yu, G. Y. Huang, M. Larsson, P. Caroff,and H. Q. Xu, Nano Lett. \textbf{12}, 6414 (2012).

\bibitem{InSb7}
H.  O.  H.  Churchill,  V.  Fatemi,  K.  Grove-Rasmussen, M.  T.  Deng,  P.  Caroff,  H.  Q.  Xu,  and  C.  M.  Marcus,Phys. Rev. B \textbf{87}, 241401(R) (2013).


\bibitem{dasS}
J. D. Sau, S. Tewari, and S. Das Sarma, \textit{Experimental and materials considerations for the topological superconducting state in electron- and hole-doped semiconductors: Searching for non-Abelian Majorana modes in 1D nanowires and 2D heterostructures}, Phys. Rev. B, \textbf{85}, 064512 (2012).

 \bibitem{soc}
Qing-feng Sun, Jian Wang and Hong Guo, \textit{Quantum transport theory for nanostructures with Rashba spin-orbital interaction}, Phys. Rev. B \textbf{71}, 165310 (2005).

\bibitem{theorySupport}
P. Szumniak, D. Chevallier, D. Loss, and J. Klinovaja, \textit{Spin and charge signatures of topological superconductivity in Rashba nanowires},
Phys. Rev. B, \textbf{96}, 041401(R) (2017).



\bibitem{pwave1}
M. T. Deng, S. Vaitiekenas, E. B. Hansen, J. Danon, M. Leijnse, K. Flensberg,
J. Nygard, P. Krogstrup, C. M. Marcus, \textit{Majorana bound state in a coupled quantum-dot hybrid-nanowire system},
Science
\textbf{354}, 1557 (2016).

\bibitem{pwave2}
E. Prada, P. San-Jose 2, M. W. A. de Moor, A. Geresdi,
E. J. H. Lee, J. Klinovaja, D. Loss, J. Nygard, R. Aguado
and L. P. Kouwenhoven,
\textit{From Andreev to Majorana bound states in hybrid superconductor-semiconductor nanowires},
Nature Reviews Physics, \textbf{2}, 575-594 (2020).

\bibitem{pwave3}
J. Alicea, Y. Oreg, G. Refael, F. V. Oppen, and M. P. A. Fisher, \textit{Non-Abelianstatistics and topological quantum information processing in 1D wire networks}, 
Nature Phys. \textbf{7}, 412-417 (2011).





\bibitem{w3}
B. A. Bernevig and T. L. Hughes, \textit{Topological Insulators
and Topological Superconductors}, (Princeton University Press,
Princeton, NJ, 2013).

\bibitem{w1}
M. Sato, Y. Tanaka, K. Yada, and T. Yokoyama,
\textit{Topology of Andreev bound states with flat dispersion},
Phys. Rev. B, \textbf{83}, 224511 (2011).

\bibitem{w2}
B. Beri,
\textit{Topologically stable gapless phases of time-reversal-invariant superconductors},
Phys. Rev. B, \textbf{81}, 134515 (2010).


\bibitem{w4}
B. Huang, X. Yang, N. Xu, and M. Gong, \textit{Type-I and type-II topological nodal superconductors withs-wave interaction}, Phys. Rev. B, \textbf{97} 045412 (2018).


\end{thebibliography}
\end{document}